# Microwave surface impedance anisotropy of $YBa_2Cu_3O_x$ single crystals with different oxygen content

Yuri A. Nefyodov [1], Mikhail R. Trunin

*Institute of Solid State Physics, Russian Academy of Sciences, 142432 Chernogolovka, Russia*

**Abstract**

The linear microwave response of ultra high-quality $YBa_2Cu_3O_x$ single crystals grown in $BaZrO_3$ crucibles is measured at 9.4 GHz in rf magnetic fields parallel and perpendicular to the $ab$-plane in the temperature range $5 \leq T \leq 200$ K. Having found the analytic solution for the magnetic field distribution on the sample surface we determine both the surface impedance $Z_{ab} = R_{ab} + iX_{ab}$ in the $ab$-plane and $Z_c = R_c + iX_c$ along $c$-axis of the crystals. For the first time the evolution of the $Z_{ab}(T)$ and $Z_c(T)$ dependences on the same sample and in a wide range of oxygen content is obtained. For $x = 6.95$ (optimum oxygen content) the temperature dependence of the imaginary part $\sigma_c''(T)/\sigma_c''(0)$ of the $c$-axis conductivity is found to be strikingly similar to that of $\sigma_{ab}''(T)/\sigma_{ab}''(0)$ and becomes more convex with $x$ lowering.

In the scope of high-$T_c$ superconductors, the problem to find out the mechanisms of quasiparticles transport along crystallographic axes of these anisotropic compounds remains unresolved. In contrast to well known linear low-temperature dependences of the penetration depth $\lambda_{ab}(T)$ and surface resistance $R_{ab}(T)$ in the $ab$-plane of single crystals, their $c$-axis microwave properties are scarce [1]. Previous surface impedance measurements [2–5] of $YBa_2Cu_3O_x$ (YBCO) were confined to the cases of the optimum doping level ($x \approx 6.95, T_c \approx 90$ K) and the so-called 60 K-phase ($x \approx 6.5$). However, even in the first case, which is the most studied one, the $c$-axis experiments gave controversial results. In particular, both linear [3,5] and quadratic [4] $\lambda_c(T)$ dependences were observed at $T < T_c/3$. The aim of this paper is to demonstrate the evolution of microwave surface impedance anisotropy of one and the same YBCO crystal in a wide range of oxygen content.

We calculated the field distribution on the surface of the sample with dimensions $b \gg a > c$ placed in microwave magnetic field $\mathbf{H}_\omega \| \mathbf{c}$. It enables us to calculate sample geometrical factor and, hence, obtain $Z_{ab}(T)$ dependence from the measurements at $\mathbf{H}_\omega \| \mathbf{c}$. Knowing the value of $Z_{ab}(T)$ we can get the $c$-axis surface impedance components from the quantities measured for the same sample at $\mathbf{H}_\omega \| \mathbf{b}$. For this we use a procedure proposed in Ref. [6]. It takes correctly account of the size effect which becomes apparent at $T > 0.9 T_c$. In more detail all these calculations will be given elsewhere [7]. The technique discussed has two main advantages: (i) the possibility to measure surface impedance anisotropy in both normal and superconducting states, which enables to obtain not only the change of penetration depths but also their absolute values and (ii) the ability to obtain the evolution of the surface impedance tensor with doping level on the same sample. With the aim of this technique we present here the results of high precision measurements of surface impedance tensor of YBCO crystal at a frequency of 9.4 GHz.

High quality YBCO single crystals were grown using $BaZrO_3$ crucibles [7]. The dynamic susceptibility measurements showed $T_c \simeq 92$K and the width of the transition less than 0.1 K at 100 KHz. After successive anneals in the air at 520, 550, 600 and 720°C transition

---

[1] Corresponding author. E-mail: nefyodov@issp.ac.ru

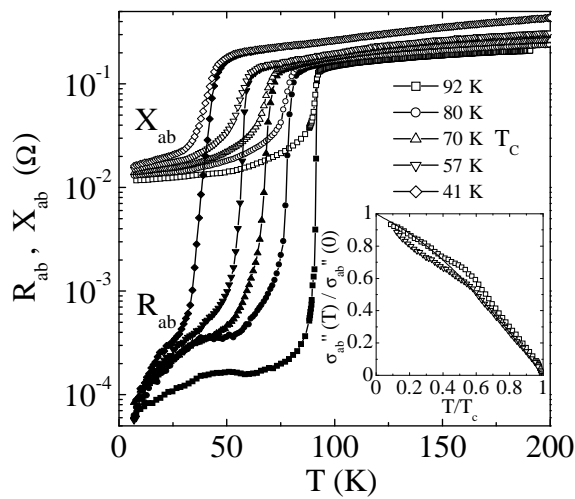 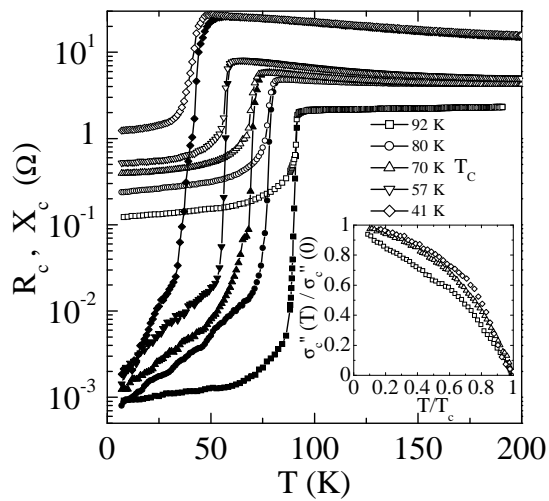

Fig. 1. Real $R_{ab}(T)$ (solid symbols) and imaginary $X_{ab}(T)$ (open symbols) parts of the $ab$-plane surface impedance of $YBa_2Cu_3O_x$ single crystal. Inset shows $\sigma''_{ab}(T)/\sigma''_{ab}(0)$ dependences for the cases of $T_c = 92$ and 57K. $\sigma''(T)/\sigma''(0)$ for the case of clean $d$-wave is shown by solid line.

Fig. 2. Real $R_c(T)$ (solid symbols) and imaginary $X_c(T)$ (open symbols) parts of the $c$-axis surface impedance of $YBa_2Cu_3O_x$ single crystal. Inset shows $\sigma''_c(T)/\sigma''_c(0)$ dependences for the cases of $T_c = 92$, 70 and 41K.

temperature reached 80, 70, 57 and 41 K respectively without significant transition broadening.

The surface impedance components $R_{ab}(T)$ and $X_{ab}(T)$ in the $ab$-plane of YBCO single crystal are shown in Fig. 1. In the normal state we have $R_{ab}(T) = X_{ab}(T)$ which implies the validity of the normal skin-effect condition. The value of residual losses $R_{ab}(T \to 0)$ does not exceed 40 $\mu\Omega$. In the case of optimum oxygen content $R_{ab}(T)$ has a broad peak at $T \sim T_c/2$ which vanishes with $x$ lowering. Both $R_{ab}(T)$ and $\lambda_{ab}(T) = X_{ab}(T)/\omega\mu_0$ dependences are linear at $T < T_c/2$. The value of the penetration depth $\lambda_{ab}(0)$ changes from 0.15 to 0.2 $\mu m$ with $x$ lowering. Being normalized by its value at zero temperature the imaginary part $\sigma''_{ab}$ of the complex conductivity exhibits temperature dependence which is very close to the one for clean superconductor with $d_{x^2-y^2}$-wave symmetry of the order parameter (solid line in the inset in Fig. 1) and does not change significantly with $x$. For the cases of optimal doping and 60 K-phase $\sigma''_{ab}(T)/\sigma''_{ab}(0)$ dependences are shown in the inset in Fig. 1.

In Fig. 2 we demonstrate temperature dependences of the $c$-axis impedance components $R_c$ and $X_c$. For optimal doping level both the components are linear at $T < T_c/2$ and $\sigma''_c(T)/\sigma''_c(0)$ curve is almost identical to $\sigma''_{ab}(T)/\sigma''_{ab}(0)$. Lowering of the oxygen content leads to the appearance of power law terms in both $R_c(T)$ and $X_c(T)$. In particular, $\Delta X_c(T) = \omega\mu_0\Delta\lambda_c(T)$ curve may be well fitted by $T^2$ dependence at $T < 25K$ in the case of $T_c = 41K$. Extrapolation to zero temperature gives the values of $\lambda_c(0)$ approximately 1.5, 3, 5, 7 and 16 $\mu m$ for $T_c = 92$, 80, 70, 57 and 41 K respectively. The $ac$-susceptibility measurements of $\lambda_c(0)$ confirmed the microwave ones within the accuracy of the former.

In conclusion, we have presented for the first time the evolution of both $Z_{ab}$ and $Z_c$ temperature dependences of $YBa_2Cu_3O_x$ single crystals in a wide range of oxygen content $x$. At optimal doping level all the surface impedance tensor components proved to be linear at $T < T_c/2$. While $R_{ab}(T)$ and $X_{ab}(T)$ curves remain linear at low temperatures with $x$ lowering, both $R_c(T)$ and $X_c(T)$ dependences become curved that reflects the reduction of the coupling between $CuO_2$ layers.

### Acknowledgements

This work was supported by RFBR grants 00-02-17053, 02-02-06578, 02-02-08004. M.R.T. thanks Russian Science Support Foundation.